\documentclass[prl,preprint,groupedaddress]{revtex4-2}

\usepackage{amsmath}
\usepackage{amssymb}
\usepackage{graphicx}
\usepackage{mathrsfs}
\usepackage{lineno}
\usepackage{bbm}
 
\begin{document}

\title{Stochastic fluctuations in protein interaction networks are nearly Poissonian.}

\author{Jaroslav Albert\\
jaroslav.albert@ronininstitute.org}
\affiliation{Ronin Institute}

\begin{abstract}
Gene regulatory networks are comprised of biochemical reactions, which are inherently stochastic. Each reaction channel contributes to this stochasticity
in different measure. In this paper we study the stochastic dynamics of protein interaction networks (PIN) that are made up of monomers and dimers. The network is defined by the dimers, which are
formed by hybridizing two monomers. The size of a PIN was defined as the number of monomers that interacts with at least one other monomer (including itself). We generated 4200 random PIN of sizes between 2 and 8 (600 per size) and simulated via the Gillespie algorithm the stochastic evolution of copy numbers of all monomers and dimers until they reached a steady state. The simulations revealed
that the Fano factors of both monomers and dimers in all networks and for all time points were close to one, either from below or above. Only 10$\%$ of Fano factors for monomers were above 1.3 and 10$\%$ of Fano factors for dimers were above 1.17, with 5.54 and 3.47 as the maximum value recorded for monomers and dimers, respectively. These findings suggests that PIN in real biological setting contribute to the overall stochastic noise that is close to Poisson. Our results also show a correlation between stochastic noise, network size and network connectivity: for monomers, the Fano factors tend towards 1 from above, while the Fano factors for dimers tend towards 1 from below. For monomers, this tendency is amplified with increased network connectivity. 
\end{abstract}

\maketitle

\section{Introduction}

Gene regulatory networks (GRN) give rise to an output of molecular products that is inherently stochastic, which leads to a cell-to-cell variability in molecular copy numbers. 
The source of this stochasticity comes from every type of biochemical reaction occurring in a cell; however, not all reactions contribute to the observable stochastic fluctuations of molecular copy numbers equally. For example, among the four basic processes that lead to a functioning protein -- transcription, translation, and degradation of mRNA and protein -- transcription has been shown to be the main source
of stochasticity, mainly due to an effect called transcriptional bursting \cite{Fujita, Cao, Bokes, Raj, Kaern}. Transcriptional bursting occurs when a gene promoter switches randomly between two (or more) promoter states: one that effectuates a high transcription rate and the other a low transcription rate. This switching leads to periods of low and high translational activity, the latter of which appears as a burst in protein copy number. There also exist translational bursts \cite{Bokes2, Lin, Kuwahara, Raj, Friedman, Ozbudak}. These occur when the mRNA copy numbers are low and the translation rate per mRNA is high. Beside the aforementioned four basic processes, there are other biochemical reactions that contribute to stochasticity, such as the formation and dissociation of dimers. The contribution of these processes to the overall copy number fluctuations of mRNA and protein has been studied to a lesser extent. At a single gene level, computational studies have been done and revealed that the presence of multimeric reaction channels tends to reduce copy number fluctuations of monomers \cite{Rooman, Pucci, Pucci2}. However, computational studies on large multimeric networks and their effects on overall noise have not been reported.

In this paper, we performed stochastic simulations of networks consisting of monomers and dimers. The translation rate for all monomers was assumed to be constant; i. e. both the average and fluctuations of mRNA copy numbers were treated as constant. This allowed us to remove the influence of mRNA on the stochastic dynamics of monomers and dimers.
The network size -- number of species of monomers -- ranged from 2 to 8. For each network size, we simulated 600 networks and computed the Fano factor for each species of monomer and dimer, for time increments of one (in minutes) as the copy numbers evolved from zero to their steady states.
Our results show the Fano factors to be on average very close to 1. These results are significant for two reasons: 1) they establish the mRNA as the primary regulator of stochastic noise;
and 2) they validate the class of hybrid stochastic simulation algorithms in which the copy numbers of monomers and dimers are assumed to be deterministic \cite{Albert}.

\section{Materials and Methods}

We employed the Gillespie algorithm (GA) \cite{Gillespie} to simulate the protein networks.
Each protein network was made up of these reactions:
\begin{eqnarray}\label{reactions}
&&1.\,\,\,\,\,\,\emptyset\xrightarrow{\makebox[1cm]{$K_i$}}n_i\nonumber\\
&&2.\,\,\,\,\,\,n_i\xrightarrow{\makebox[1cm]{$q_i$}}\emptyset,\nonumber\\
&&3.\,\,\,\,\,\,n_i+n_j\xrightarrow{\makebox[1cm]{$\alpha_{ij}$}}s_{ij}\nonumber\\
&&4.\,\,\,\,\,\,s_{ij}\xrightarrow{\makebox[1cm]{$\beta_{ij}$}}n_i+n_j\nonumber\\
&&5.\,\,\,\,\,\,s_{ij}\xrightarrow{\makebox[1cm]{$d_{ij}$}}\emptyset,\nonumber\\
\end{eqnarray}
where $i$ and $j$ label the genes in the network. The changes in copy numbers these reactions bring about are as follows: 1. monomer copy number of gene $i$, $n_i$,
increases by one with propensity $K_i$; 2. due to degradation, $n_i$ decreases by one with propensity $q_in_i$; 3. monomer of gene $i$ combines with a monomer of gene $j$
to form a dimer, thereby decreasing $n_i$ and $n_j$ by one and increasing dimer copy number $s_{ij}$ by 1 
with propensity $\alpha_{ij}n_i(n_j-\delta_{ij})/(1+\delta_{ij})$ (if $i=j$, $n_i$ decreases by 2); 4. dimer consisting of monomers $i$ and $j$
dissociates into monomers $i$ and $j$, thereby increasing $n_i$ and $n_j$ by one and decreasing $s_{ij}$ by 1 
with propensity $\beta_{ij}s_{ij}$ (if $i=j$, $n_i$ increases by 2); and 5. due to degradation $s_{ij}$ decreases by 1.
We define a protein network as a set of nodes and edges, where monomer $i$ is a node and the interaction between monomer $i$ and monomer $j$, whose strength is
given by $\alpha_{ij}$, is an edge. If $\alpha_{ij}$ is zero, there is no interaction -- no edge -- between monomers $i$ and $j$.
The network size $N$ is determined by the number of monomer species, while network connectivity
is given by the total number of edges. We also defined a connection probability, $c$, and a sampling function
$C_{ij}(c)=\theta(c-\epsilon_{ij})$, where $\epsilon_{ij}$
is a random number between 0 and 1, and $\theta(x)$ is the Heaviside function. If $C_{ij}=1$, monomer $i$ interacts with monomer $j$; otherwise where is no interaction. 
If $c$ is high, $C_{ij}$ is likely to be one, which leads to a highly connected network. A low $c$
generates a sparse network.

We have generated 4200 networks: 200 networks for each $N=[2,8]$ and $c=\{0.1,0.5,0.8\}$. For every $N$ and $c$,
system parameters $K_i$, $b_i$, $\alpha_{ij}$, $\beta_{ij}$, $d_{ij}$, were selected using a random number generator with the following ranges:
$K_i=[10,1000]$min$^{-1}$, $b_i=[5, 500]\times10^{-3}$min$^{-1}$, $d_{ij} = (b_i + b_j)/2\times[0.2, 1]$min$^{-1}$, $\beta_{ij} = [1,50]\times10^{-3}$min$^{-1}$, 
$\alpha_{ij} = C_{ij}(c)\times[1,10]\times10^{-4}$min$^{-1}$. Also, since $s_{ij}$ is the same species as $s_{ji}$, only the elements $\alpha_{i\leq j}$ are needed; all others were set to zero.
Same applies to the matrices $\beta_{ij}$ and $d_{ij}$, as well as the dimer copy numbers $s_{ij}$. For convenience, the copy numbers of the dimer species that are actually formed via reaction 3 in Eq. (\ref{reactions}) will be denoted by a single upper index: $s^k$, such that $k=1,...,M$, where $M$ is the total number of dimer species.  
A set of mass action kinetics equations (MAKE) were integrated to obtain steady state values of copy numbers for all monomer and dimer species;
if the minimum copy number was larger than 100, the network topology was saved. This cutoff is arbitrary and will be mentioned in the discussion section. 

The time at which steady state $T_{ss}$ was reached was done like this:
The MAKE were integrated between the time $t=0$ and $t=m$, where the integer $m$ was incremented by 1. 
Each time $m$ was increased, the slopes $(dn_i/dt)|_{t=m}$ and $(ds^k/dt)|_{t=m}$ were computed and
added to the set $Mn_i=[(dn_i/dt)|_{t=1},...,(dn_i/dt)|_{t=m-1}]$ 
and $Ms^k=[(ds^k/dt)|_{t=1},...,(ds^k/dt)|_{t=m-1}]$, respectively.
The slopes were renormalized with respect to the maxima $\text{max}[Mn_i]$ and $\text{max}[Ms^k]$:
${\tilde M}n_i=[(dn_i/dt)|_{t=1},...,(dn_i/dt)|_{t=m}]/\text{max}[Mn_i]$ and 
${\tilde M}s^k=[(ds^k/dt)|_{t=1},...,(ds^k/dt)|_{t=m}]/\text{max}[Ms^k]$.
When $\text{max}[{\tilde M}n_1,...,{\tilde M}n_N,{\tilde M}s^1,...,{\tilde M}s^M]<0.1$, $T_{ss}$ was set to $m$.
The reason for this cutoff was to ensure that the system is close to a steady state, while minimizing the running time of the GA.
All 4200 protein networks were simulated by the GA with a sample size of 100. The initial values for all variables were 0; the algorithm stopped when
the accumulated sampled time reached $T_{ss}$.

\section{results and discussion}

The results of our simulations can be seen in figures 1, 2 and 3 below.

Figure 1 shows a random example from the 200 networks with $c=0.8$ and $N=5$ of the average and standard deviation of monomers and
dimers, and the corresponding Fano factors for the time points 0 to $T_{ss}$. The fluctuations, most clearly visible for the Fano factors,
come from the relatively small sample size (100). 

Figure 2 shows the averages and standard deviations for Fano factors for each $N$ and $c$. The sample from which these quantities were
computed was constructed by listing the Fano factors of all species of monomers in all 200 networks at all time points. Same was done for dimers.
The graphs show that the average Fano factor and its standard deviation tend to decrease as the network size increases and as the connectivity of the
network increases. For dimers, with increasing $N$, the average tends to increase, while the standard deviation tends to decrease. This trend is only slightly enhanced by
higher connectivity. In all cases, however, the Fano factors are close to one, either from above or below. 

In Figure 3 a) we show a distribution of Fano
factors for all $N=2,...,8$ and $c=0.1,0.5,0.8$ at all time points, i. e. all available data points, for monomers (in blue) and dimers (in purple). Fig. 3b) shows the reverse cumulative distribution. The upper dashed line shows that 50$\%$ of Fano factors are above 1.02, for monomers, and 0.97, for dimers. The lower dashed line indicates that Fano factors higher than 1.3, for monomers, and 1.17 for dimers are
only in the 10$^{\text{th}}$ percentile.

\begin{figure}
\centering
\includegraphics[trim=0 0 0 1.0cm, height=0.8\textheight]{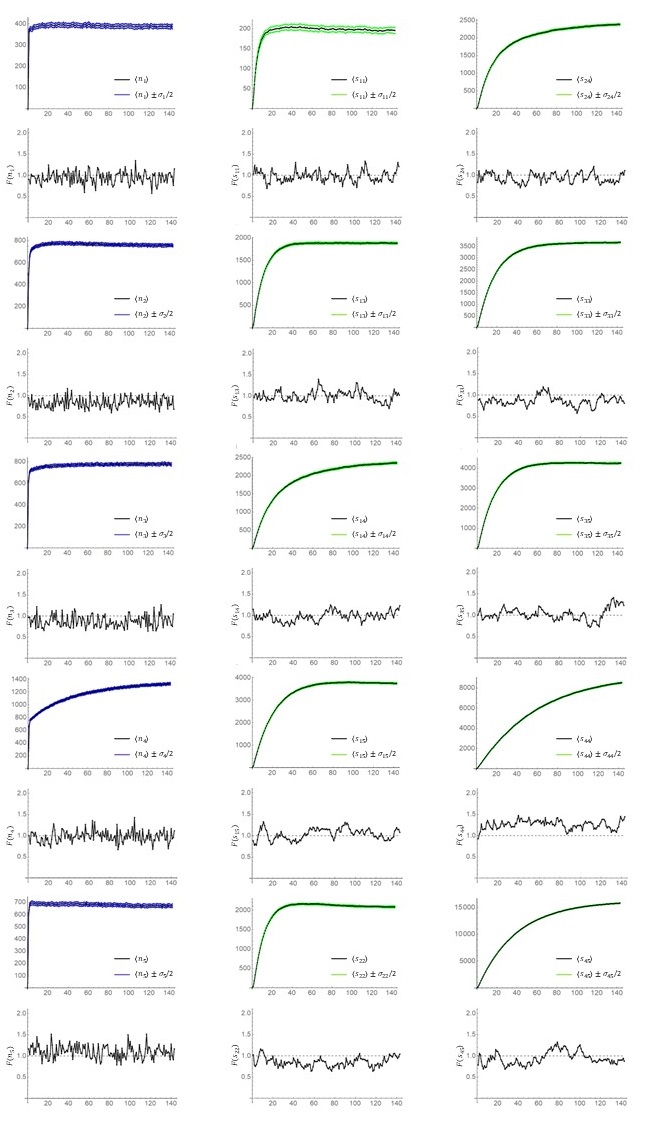}
\caption{\small{An example for $N=5$ and $c=0.8$ of the average and standard deviation of monomers and
dimers (odd rows), and the corresponding Fano factors for the time points 0 to $T_{ss}$ (even rows). The averages for monomers (dimers) appear in black, sandwiched between the
blue (green) curves that represent the standard deviation.}}
\end{figure}

\begin{figure}
\centering
\includegraphics[trim=0 0 0 1.0cm, height=0.27\textheight]{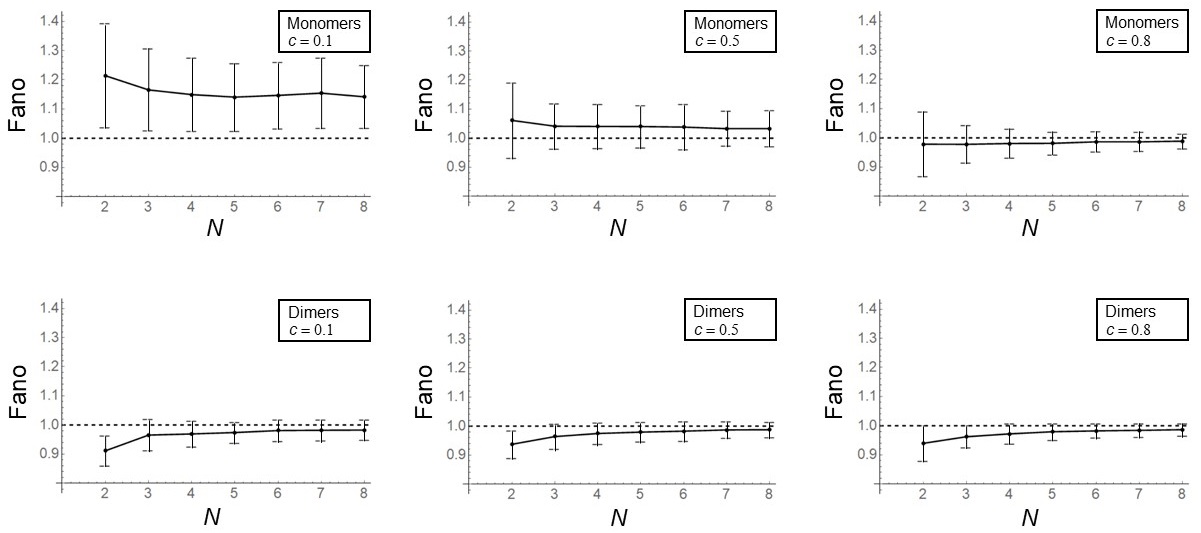}
\caption{Averages and standard deviations for Fano factors for each $N$ and $c$. Each graph was constructed from a sample
containing all monomers (top row) and dimers (bottom row) at all recorded time points.}
\end{figure}

\begin{figure}
\centering
\includegraphics[trim=0 0 0 1.0cm, height=0.2\textheight]{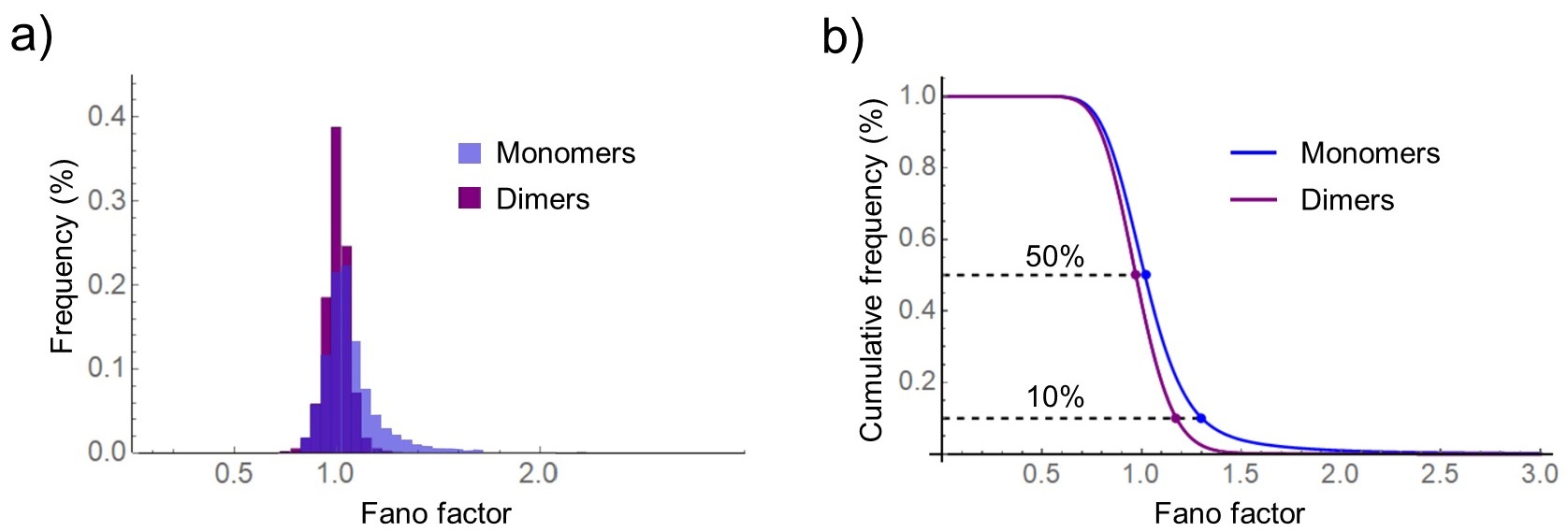}
\caption{a) Distribution of Fano
factors for all recorded data points, for monomers (in blue) and dimers (in purple); b) Reverse cumulative distribution of Fano
factors for all recorded data points, for monomers (in blue) and dimers (in purple).}
\end{figure}

This work, while serving as a useful indicator, is not a comprehensive study of protein networks present in real biological cells; such networks are far more complex and involve many additional reaction pathways not considered in our models. For example, our study does not take into account the post-translation modification processes or the role of signaling pathways which can render a monomer or a
dimer (in)effective, thus placing it in a different species category. Also, real protein networks are typically much larger than eight monomers. Therefore, one cannot be sure whether our results apply to
networks with $N\gg 8$.  However, the purpose of this study was only to establish a trend, which does indicate a tendency towards Poisson statistics as $N$ increases: the average Fano factors approach 1, while the standard deviations from this average diminish with $N$. This tapering of standard deviation would likely be even more pronounced if our ensemble had been larger. As Figure 1 has shown, the fluctuations of Fano factors due to our relatively small ensemble of 100 are quite prominent. 

In conclusion, we have demonstrated that protein networks comprising of monomers and dimers, and in the absence of any stochastic factors extrinsic to it, such as fluctuations in mRNA copy numbers, are nearly Poissonian. This finding is significant when trying to assign contributions from various reactions of a real GRN to the overall level of stochastic noise. Since Poisson noise is generally considered to be
low, at least in the context of GRN, its contribution is not a major one. This information can be of great value to researchers in the field of stochastic modeling, as well as in
the increasingly prevalent experimental investigations of stochasticity in GRN. Our results also help to validate a recently proposed hybrid stochastic simulation algorithm \cite{Albert}, which, in one of its
versions, relied on the statistics of the kind of protein networks we studied here to be close to Poisson.

\end{document}